# A Data-Driven Digital Twin Network Architecture in the Industrial Internet of Things (IIoT) Applications

Abubakar Isah[1], Hyeju Shin[1], Ibrahim Aliyu[1], Sangwon Oh[1], Sangjoon Lee[2], Jaehyung Park[1], Minsoo Hahn[3], Jinsul Kim[*]

[1]Department of ICT Convergence System Engineering, Chonnam National University, Gwangju, Korea.
[2]Interdisciplinary Program of Digital Future Convergence Service, Chonnam National University, Gwangju, Korea.
[3]Astana IT University, Astana, Kazakhastan.

abubakarisah@jnu.ac.kr[1], sinhye102@gmail.com[1], aliyu@ieee.org[1], osw0788@gmail.com[1], s-lee@jnu.ac.kr[2], hyeoung@jnu.ac.kr[1], m.hahn@astanait.edu.kz[3], jsworld@jnu.ac.kr[*]

*Abstract*

*A new network named the "Digital Twin Network" (DTN) uses the "Digital Twin" (DT) technology to produce virtual twins of real things. The network load and size continue to grow as a result of the development of 5G, the Internet of Things, and cloud computing technology as well as the advent of new network services. As a result, network operation and maintenance are becoming more difficult. A digital twin connects the real and digital worlds, exchanging data in both directions and revealing information about the progression of a network process. The framework of the Industrial Internet of Things, data processing, and digital twin network is taken into consideration in this article as a key aspect. This paper proposed a data-driven digital twin network architecture, that comprises the physical network layer (PNL), the digital twin layer(DTL), the application layer (AL), and what those layers encompass and beyond. Also, we presented DTN data types and protocols to be used for data integration.*

**Keywords:** *Digital Twin Network, Industrial Internet of Things, 5G*

## 1. Introduction

Digital Twin (DT) is a technological approach that integrates artificial intelligence (AI), the Internet of Things (IoT), and virtual and augmented reality (VR/AR) to form models or simulations of physical objects, systems, or procedures.

The Internet of Things (IoT) providers are aware of how desirable the Industrial Internet of Things (IIoT) business is compared to the IoT market. Thus, Industrial (IoT) market [1] vendors are bringing device and information processing agility [2]. Digital Twin is a synchronized, real-time virtual depiction of a process, environment, or product [3]. Therefore, Industrial control system change must be properly researched and analyzed before being implemented to minimize environmental and safety hazards since industrial data processing regulates the physical processes [4]. However, due to the less severe repercussions of failure, corporate data processing does not require the same strict controls.

In addition, the Industrial IoT mandates monitoring the whole production process. Digital Twin analyzes data and offers a virtual environment to evaluate the manufacturing process's functioning before it is used in the real world. [5]

Furthermore, in order to actualize the concept of a cyber-physical continuum between the physical world and its digital representation, DTNs will serve as a fundamental 6G technology [6]. One of the most promising technologies that can help 6G communication systems achieve their technical and commercial goals is DT technology [7].

In this study, we aim to present a data-driven digital twin network architecture in IIoT applications which consists of three layers. The Digital Twin Network comprises three layers: The physical network layer, the twin network layer, the application layer, and their interfaces. The layers consist of a group of data domains, model domain, twin management, and a detailed explanation of those three layers. We also presented DTN data types, descriptions, and protocols to consider. The order of this paper was as follows: The related research was covered in Section 2. The digital twin network proposed data-driven architecture was created in Section 3. In Section 4, the IIoT data type idea and related protocols were also introduced. In Section 5, the conclusion and future work were covered.

## 2. Related Work

Many research studies have been conducted to investigate the concept of DTN [8] and propose various frameworks [9], and networking services in the context of smart factories [6], [7]. These studies aim to advance the understanding and implementation of DTN in industrial settings.

The industrial revolution concept, [2] presented a framework and was viewed as a key role in the development of new industry capabilities for industrial production control systems in IIoT. A method for dynamically reconfiguring production systems, including their layout, the process variables, and operation times of various assets, in response to shifting customer or market needs by fusing digital twins with modular artificial intelligence algorithms [10].

However, to meet the technical, governmental, and commercial demands of the network communication service users, a fresh set of difficult requirements and more stringent key performance indicators must be taken into account [7], a novel architecture must be designed, and special enabling technologies must be created. Our focus on the framework design will be data-driven which must be produced from the gathered information to achieve the objectives of network monitoring in IIoT applications.

## 3. Digital Twin Network Architecture

The architecture consists of two closed loops: an internal closed loop that supports simulation and iterative optimization based on a service mapping model, and an exterior closed loop that regulates network applications based on all-three-layer architecture.

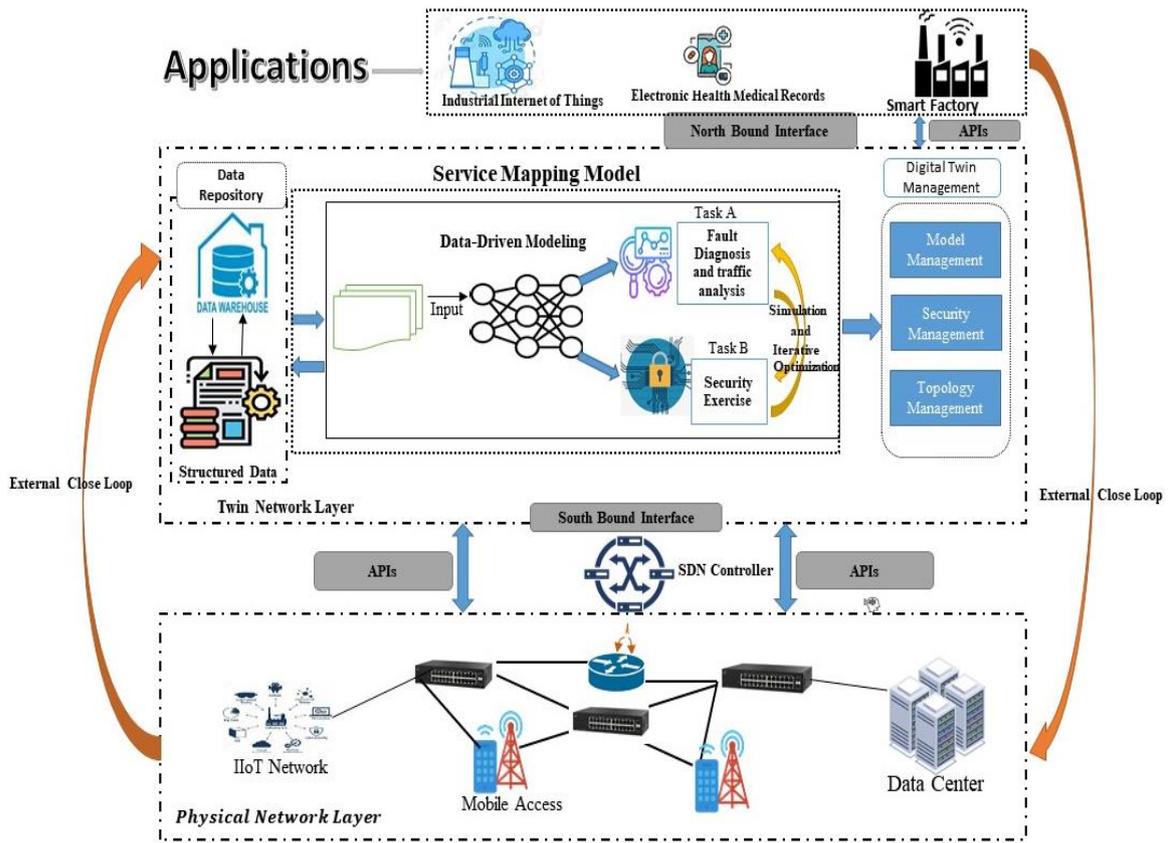

**Figure 1:** The Data-Driven Digital Twin Network Architecture.

## 3.1    The Application Layer.

The network application layer serves as an interface between the digital twin network and outside applications [11], giving these programs a way to take advantage of the ability of the digital twin network for different uses. This layer makes it possible for external applications to utilize the data and services offered by the digital twin network for sophisticated use cases like real-time decision-making and predictive analytics. The network application layer is crucial in maximizing the potential of digital twin technology because it enables a seamless connection between the digital twin network and external applications.

## 3.2    Twin Network Layer.

The twin network layer, which manages and organizes the network's data, services, and applications, is an essential component of the digital twin network. This layer lies between the network application layer and the physical network layer in DTN [11]. It consists of three subsystems: twin network management (model topology, security management, etc.), service mapping model (functional and basic model, and the data repository (data sharing and data services).

## 3.3    Physical Network Layer.

Through the twin network's South Bound Interface (SBI), the physical network layer makes it possible for physical network devices and the network twin to exchange network data and control information [12]. This layer comprises a range of physical network types, including IIoT networks, data center, and mobile access networks, and transport networks.

## 4. Data Type in Digital Twin Network

The process of merging data from many sources and making it available for analysis, reporting, and decision-making processes is known as data integration [12]. Several tools can be used to collect, transmit, and manage real-time data in the Digital Twin Network. Figure 3: Give information on the most popular data collection protocols in Industrial Internet of Things (IIoT) applications.

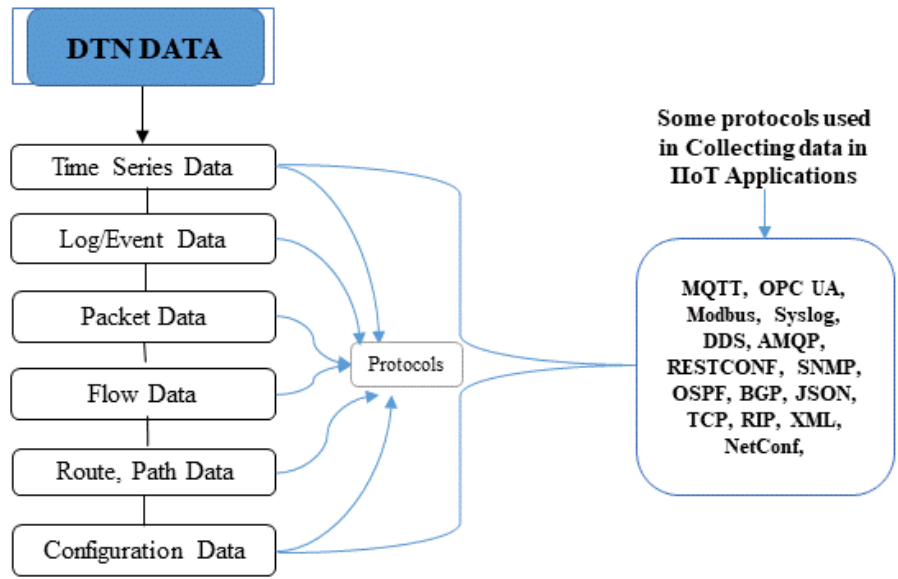

**Figure 2**: Data and Common Protocols in IIoT Applications

Description of the aforementioned data in DTN: To explain more about the

**Table 1.** Digital Twin Network Data.

| Data Types | Protocols | Description |
|---|---|---|
| Time Series Data | MQTT, OPC UA, REST, CoAP, and AMQP | Information from sensors or performance metrics obtained data in real-time. |
| Event Data | AMQP, MQTT, CoAP, Syslog, DDS, and JMS | Devices, applications, and systems that generate logs that capture events, failures, and activities. |
| Packet Data | MQTT, CoAP, OPC UA, DDS, and AMQP | Refers to the individual packets of data that are transmitted over a network |
| Flow Data | Modbus/TCP, OPC UA, MQTT, CoAP, and DDS | Flow data is a term for cluster data that describes the properties of network flows, which are collections of packets with similar properties. |
| Route Path Data | OSPF, BGP, IS-IS, RIP, and SNMP | Refers to the information needed for routing and forwarding decisions in a network. |
| Configuration Data | SNMP, NetConf, RESTCONF, YANG, JSON, and XML | The data include settings parameters and configurations that are used to configure devices or systems in DTN |

There are protocols or techniques for data collection in IIoT applications that may exist based on the particular use case and the devices being monitored. Some protocols may also be used to gather various kinds of data. Any protocol used to gather IIoT data should be carefully evaluated in terms of security and scalability.

## 5. Conclusion

In this paper, a novel data-driven DTN architecture has been proposed for Industrial Internet of Things applications. The framework consists of 3-layer architecture and its interfaces. The physical network layer interfaces with the twin network layer through the SIB in which software-defined networks play an important role in securing the communication and data exchange. The application layer communicates with the twin network layer through the NBI and a dual closed loop was present. Moreover, we also proposed so data type and their corresponding protocols for generating such data.

In the future, there is a need to identify the tools to be used for collecting data based on the case study, data streaming platforms, integration and interoperability of such complex data. In that case, security, and scalability, and devices need to be considered while collecting such high dimensional data while deploying such system.

## Acknowledgment

This work was supported by Electronics and Telecommunications Research Institute (ETRI) grant funded by the Korean government [22ZR1100, A study of hyper-connected thinking Internet technology by autonomous connecting, controlling and evolving ways] and supported by the MSIT(Ministry of Science and ICT), Korea, under the Innovative Human Resource Development for Local Intellectualization support program(IITP-2023-RS-2022-00156287) supervised by the IITP(Institute for Information & communications Technology Planning & Evaluation)

6   10th ICAEIC-2023## References

[1]  N. Khilji, A. Isah, and K. Usman, "Prospective: Smart Cities with the Internet of Things," no. March, pp. 5–6, 2016, [Online]. Available: http://sdtechnocrates.com/ETEBMS2016/html/papers/ETEBMS-2016_ENG-CS4.pdf

[2]  M. B. Sergeeva, V. V. Voskobovich, and A. M. Kukharenko, "Data Processing in Industrial Internet of Things (IIoT) Applications : Industrial Agility," *2022 Wave Electron. its Appl. Inf. Telecommun. Syst. WECONF 2022 - Conf. Proc.*, pp. 4–8, 2022, doi: 10.1109/WECONF55058.2022.9803390.

[3]  W. Purcell and T. Neubauer, "Digital Twins in Agriculture: A State-of-the-art review," *Smart Agric. Technol.*, vol. 3, no. July 2022, p. 100094, 2023, doi: 10.1016/j.atech.2022.100094.

[4]  A. R. Al-Ali, R. Gupta, T. Z. Batool, T. Landolsi, F. Aloul, and A. Al Nabulsi, "Digital twin conceptual model within the context of internet of things," *Futur. Internet*, vol. 12, no. 10, pp. 1–15, 2020, doi: 10.3390/fi12100163.

[5]  C. Metallidou, K. E. Psannis, D. D. Vergados, and M. Dossis, "Digital Twin and Industrial Internet of Things Architecture to Reduce Carbon Emissions," *2022 4th Int. Conf. Comput. Commun. Internet, ICCCI 2022*, pp. 185–189, 2022, doi: 10.1109/ICCCI55554.2022.9850248.

[6]  X. Lin, L. Kundu, C. Dick, E. Obiodu, and T. Mostak, "6G Digital Twin Networks: From Theory to Practice," pp. 1–7, 2022.

[7]  N. P. Kuruvatti, M. A. Habibi, S. Partani, B. Han, A. Fellan, and H. D. Schotten, "Empowering 6G Communication Systems With Digital Twin Technology: A Comprehensive Survey," *IEEE Access*, vol. 10, no. October, pp. 112158–112186, 2022, doi: 10.1109/ACCESS.2022.3215493.

[8]  J. Friederich, D. P. Francis, S. Lazarova-Molnar, and N. Mohamed, "A framework for data-driven digital twins for smart manufacturing," *Comput. Ind.*, vol. 136, Apr. 2022, doi: 10.1016/j.compind.2021.103586.

[9]  L. Hui, M. Wang, L. Zhang, L. Lu, and Y. Cui, "Digital Twin for Networking: A Data-driven Performance Modeling Perspective," *IEEE Netw.*, 2022, doi: 10.1109/MNET.119.2200080.

[10]  F. Mo *et al.*, "A framework for manufacturing system reconfiguration and optimization utilising digital twins and modular artificial intelligence," *Robot. Comput. Integr. Manuf.*, vol. 82, no. January, p. 102524, 2023, doi: 10.1016/j.rcim.2022.102524.

[11]   dkk 2018 ) richard oliver ( dalam Zeithml., "済無No Title No Title No Title," *Angew. Chemie Int. Ed. 6(11), 951–952.*, no. April, pp. 2013–2015, 2021.

[12]  K. Findings, "Rethink Network Monitoring for the Cloud Era," no. December, 2018.